\documentclass{article}
\usepackage[a4paper, margin=1.5cm]{geometry}
\usepackage{cite}
\usepackage[pdftex]{graphicx}
\usepackage{amsmath}
\usepackage{amsfonts}
\usepackage{array}
\usepackage[caption=false,font=footnotesize]{subfig}
\usepackage{caption}
\usepackage{multirow}
\usepackage{float}
\usepackage[table]{xcolor}
\def\url#1{}

\newcommand{\red}[1]{#1}
\newcommand{\pink}[1]{#1}
\newcommand{\pinkrow}[1]{#1}

\begin{document}

\title{A neural network for forward and inverse nonlinear Fourier transforms for fiber optic communication}

\author{Wen Qi Zhang, Terence H. Chan and Shahraam Afshar V.
\thanks{W. Q. Zhang and S. Afshar are with Laser Physics and Photonic Devices Laboratories, STEM, University of South Australia, Australia.}
\thanks{T. H. Chan is with the Institute for Telecommunications Research, University of South Australia, Australia}
\thanks{e-mail: wenqi.zhang@unisa.edu.au}
\thanks{Manuscript received April 19, 2005; revised August 26, 2015.}}

\markboth{Journal of \LaTeX\ Class Files,~Vol.~14, No.~8, August~2015}%
{Shell \MakeLowercase{\textit{et al.}}: Bare Demo of IEEEtran.cls for IEEE Journals}

\maketitle

\begin{abstract}
We propose a neural network for both forward and inverse continuous nonlinear Fourier transforms, NFT and INFT respectively. We demonstrate the network's capability to perform NFT and INFT for a random mix of NFDM-QAM signals. The network transformations (NFT and INFT) exhibit true characteristics of these transformations; they are significantly different for low and high-power input pulses. The network shows adequate accuracy with an RMSE of $5\times10^{-3}$ for forward and $3\times10^{-2}$ for inverse transforms. We further show that the trained network can be used to perform general nonlinear Fourier transforms on arbitrary pulses beyond the training pulse types.

\end{abstract}

\maketitle

\section{Introduction}
Fibre optics communication using nonlinear Fourier transform (NFT) shows the potential to break the infamous Shannon linear capacity limit caused by the nonlinear light-matter interaction in current linear Fourier transform-based optical fibre channels.~\cite{yousefi_information_2014, yousefi_information_2014-1, yousefi_information_2014-2, shannon_mathematical_1948}. The nonlinear spectrum associated with NFT consists of a continuous and a discrete part. The continuous part of the spectrum resembles the linear Fourier spectrum, for low-energy pulses, while the discrete part is related to optical solutions~\cite{yousefi_information_2014}. Preliminary experimental demonstrations show that by utilizing the continuous part of the nonlinear spectrum, NFT-based communication can indeed outperform conventional linear Fourier transform-based communications scheme~\cite{le_demonstration_2017, le_125_2017, le_high_2018}. However, one major challenge to speeding up the development of NFT-based communication systems is an easy approach to implementing NFT-based systems in hardware.

The computational complexity of NFT and its inverse, iNFT, has become the bottleneck for practical implementation~\cite{turitsyn_nonlinear_2017, lima_nonlinear_2017}. Recent attempts towards building FPGA-based NFT platforms reveal the difficulties in converting a numerical algorithm to hardware, such as the complexity in design, the limit of using only fix-point numbers, the scalability and challenge for parallel computing~\cite{vasylchenkova_fixed-point_2021}. To circumvent the challenge, an alternative approach of using neural networks in addition to advanced AI hardware has been proposed~\cite{zhang_serial_2022}, where convolutional neural networks are used to directly decode information from optical bursts. The application of neural networks for NFT can be traced back to 2018, although early works mainly focused on post-processing such as constellation classification \cite{kotlyar_combining_2020} and equalisation~\cite{kotlyar_unsupervised_2018, kotlyar_convolutional_2021}, or classification types of neural networks for solitons~\cite{jones_time-domain_2018} and symbols~\cite{zhang_direct_2021}. 

Designs of neural networks aiming at converting optical signals directly to their nonlinear spectra have been attempted recently~\cite{sedov_neural_2021, sedov_neural_2021-1}. However, these works are still limited to the direct mapping between a fixed set of input and output. There is not yet a trained neural network with the ability to work with varying input parameters, such as pulse energy, pulse width, pulse bandwidth, type of carrier wave function and modulation format. More importantly, even though some claim their network design can perform NFT, there is not yet a network that reflects the important characteristic of NFT, namely at high pulse energy, the nonlinear and linear Fourier spectra are different, while at low pulse energy, the nonlinear Fourier spectra converge towards their linear counterparts. 

This paper proposes a neural network architecture for both forward and inverse NFTs on optical signals with only continuous spectra that can address the points mentioned earlier. In addition, with proper training, our design is capable of approximating the important characteristics of NFT as mentioned above. Furthermore, we demonstrate the ability of the trained network to perform NFT and iNFT on pulses outside the scope of the training dataset.

\section{Nonlinear Fourier transform and the modulation of the continuous spectrum}
NFT, also called inverse scattering transform, is a mathematical approach that can be used to solve nonlinear Schr\"{o}dinger equation 
that describes the behaviour of optical pulses propagating in optical fibres\cite{yousefi_information_2014, zhang_correlated_2019}:
\begin{equation}
    \frac{\partial}{\partial z}q(t,z)=-i\frac{\partial^2}{\partial t^2}q(t,z)-2i\left|q(t,z)\right|^2 q(t,z),
\end{equation}
in which $q(t,z)$ is the complex optical pulse at time $t$, and propagation position $z$ along the fiber. In NFT $q(t,z)\rightarrow Q(\lambda,z)$, where $\lambda$ is the nonlinear frequency, $Q(\lambda,z)$ is the nonlinear spectrum, and the propagation of $Q(\lambda,z)$ through the fibre is given by\cite{yousefi_information_2014}
\begin{equation}
    Q(\lambda,z)=Q(\lambda,0)e^{-4i\lambda^2 z}
\end{equation}
A simple mathematical description of NFT is equivalent to solving the following differential equation~\cite{yousefi_information_2014}
\begin{equation}
	v_{t}=\left(
	\begin{array}[c]{ccc}
		- i\lambda & q(t) & \\
		-q^{*}(t) & i\lambda &
	\end{array}
	\right)v\text{,}
    \label{eq1}
\end{equation}
where $q(t)=q(t,0)$ with the initial condition:
\begin{alignat}{3}
	v(t,\lambda) &= \left(	\begin{array}{l}
                                    v_1(t,\lambda)\\
                                    v_2(t,\lambda)
                                \end{array} \right) \to \left(  \begin{array}{l}
                                                                                1 \\
                                                                                0
                                                                            \end{array} \right)e^{-j\lambda t}, & \quad t\to -\infty \label{eq:v2t_inf}
\end{alignat}
where $\lambda$ is the nonlinear frequency (also called eigenvalues), and $v_t$ is the temporal derivative of $v$. When solving the above equation, two time-invariant variables $a(\lambda)$ and $b(\lambda)$ can be defined using the following limits~\cite{yousefi_information_2014}:
\begin{equation}
\begin{array}{ccc}
    a(\lambda)=\displaystyle\lim_{t\to\infty} v_1(t,\lambda)e^{j\lambda t}, & & b(\lambda)=\displaystyle\lim_{t\to\infty} v_2(t,\lambda)e^{-j\lambda t}.
\end{array}
\label{eq:eqab}
\end{equation}
The nonlinear spectrum of the signal $q(t)$ is defined as
\begin{alignat}{2}
	Q(\lambda) &= \left\{
        \begin{array}{l}
	       b(\lambda)/a(\lambda) \quad \text{for} \quad \lambda\in\mathbb{R}\text{,}\\
	       b(\lambda)/a^\prime(\lambda) \quad \text{for} \quad \lambda\in\mathbb{C^+} \text{ and } a(\lambda)=0\text{,}
        \end{array} \right.\label{eq:Q}
\end{alignat}
where $\displaystyle a^\prime(\lambda)=\frac{\partial a(\lambda)}{\partial\lambda}$. When $\lambda$ is real, the nonlinear Fourier spectrum is continuous and when it is in the upper complex plane, $a(\lambda)=0$ and the spectrum is discrete. Discrete spectral points correspond to optical solitons. The continuous spectrum can be used, similar to the linear Fourier spectrum, to carry information using frequency division multiplexing type of modulation schemes~\cite{yousefi_information_2014}.

Nonlinear frequency division multiplexing (NFDM) is a popular modulation scheme used for NFT communication~\cite{le_125_2017, aref_modulation_2018, gemechu_comparison_2017, le_achievable_2016, derevyanko_capacity_2016, civelli_mitigating_2020}. The modulation is usually applied to the continuous spectrum directly, but modulating $b(\lambda)$ (b-modulation) has also been used~\cite{wahls_generation_2017, gui_nonlinear_2018, le_high_2018, yangzhang_400_2018, vasylchenkova_signal-dependent_2019, wahls_wiener-hopf_2019}. The modulation can often be expressed as
\begin{equation}
    s(\lambda)=\sum_n c_n w_n(\lambda)\text{,}
    \label{eq:s}
\end{equation}
where $n$ is the number of subcarriers, $c_n$ is the complex data symbols, $w_n(\lambda)$ is the carrier wave function, $s(\lambda)$ is the modulated spectrum. To adjust the pulse energy, the spectrum $Q(\lambda)$ is scaled by a factor $A$. The pulse energy $E$ can be calculated using $\displaystyle E = 1/\pi \cdot \int \ln{(1+\left|A\cdot Q(\lambda)^2\right|)} d\lambda$\cite{yousefi_information_2014}. In the case of the b-modulation scheme, $s(\lambda)$ is $b(\lambda)$ and to obtain $Q(\lambda)$, one needs to first calculate $a(\lambda)$ from $b(\lambda)$~\cite{wahls_generation_2017}. Scaling the pulse energy with b-modulation is not trivial. Hence, we focus on modulating $Q(\lambda)$ only in this work, i.e., we consider $Q(\lambda)=s(\lambda)$.

\section{The network architecture and training dataset}
To design a neural network for NFT, we examine the integral form of the solutions of Eq.~\ref{eq1}  (Eqs. 1.3.5b to 1.3.5d in \cite{ablowitz_solitons_1981}, noting that we have replaced $\phi_1$ and $\phi_2$ with $v_1$ and $v_2$, $x$ and $y$ with $t$ and $u$ to match the notations here),
\begin{eqnarray}
    v_1(t,\lambda) e^{i\lambda t} = 1 + \displaystyle\int_{-\infty}^t \left[ M(t,u,\lambda) v_1(u,\lambda)\right] e^{-i\lambda u} du, \label{eq:v1_int1} \\
    v_2(t,\lambda) e^{i\lambda t} = \displaystyle\int_{-\infty}^t \left[q^\ast(u) v_1(u,\lambda)e^{2i\lambda t}\right] e^{-i\lambda u} du, \label{eq:v1_int2}
\end{eqnarray}
where
\begin{equation}
    M(t,u,\lambda) = - q^\ast(u) \int_u^t q(v) e^{2i\lambda v} dv. \label{eq:M}
\end{equation}
\red{These solutions can be confirmed by direct substitution of Eqs. \ref{eq:v1_int1} to \ref{eq:M} into Eq. \ref{eq1}.} Four points can be drawn from these equations that guide us to design the network. (1) Using Eqs. \ref{eq:eqab}, \ref{eq:Q} and \ref{eq:v1_int1} to \ref{eq:M}, we can find \red{

\begin{eqnarray}
     M(t,u,\lambda) \to 0, 
     \\
    v_1(t,\lambda) e^{i\lambda t} \to 1, 
    \\
    v_2(t,\lambda) e^{i\lambda t} \to \displaystyle\int_{-\infty}^t \left[q^\ast(u) e^{2i\lambda t}\right] e^{-2i\lambda u} du, \label{simple}
\end{eqnarray}
in the limit of small $q$ and $q^\ast$ and hence}
\begin{equation}
Q(\lambda)=\frac{\displaystyle\lim_{t\to\infty}v_2(t,\lambda)e^{-i\lambda t}}{\displaystyle\lim_{t\to\infty}v_1(t,\lambda)e^{i\lambda t}} \to \int_{-\infty}^{\infty} q^\ast(u)e^{-2i\lambda u}du
\end{equation}
\red{for this limit}, which is the linear Fourier transform of $q(u)$ where the linear angular frequency is $2\lambda$. This shows that the nonlinear spectra of low-energy signals should converge to their linear spectra. Hence, we decided to use the linear and nonlinear spectra of the optical signal as the input and output of the network respectively. \red{In this way, the neural networks only need to learn the difference between the spectra when the signal energy is high.} (2) There are multiplications in Eqs.~\ref{eq:v1_int1},~\ref{eq:v1_int2} and \ref{eq:M} that can be approximated by convolutional layers in a network. (3) The integral in $M(t,u,\lambda)$ is a linear Fourier transform over a time-dependent interval of $u$ and $t$. To reflect this time dependence, we introduce Long-Short Term Memory (LSTM) layers into the network~\cite{hochreiter_long_1997}. LSTM networks are a kind of recurrent neural network that shows good performance at modelling long-term dependence in data. (4) the nonlinear spectra can be solved iteratively with linear spectra as the input. Therefore, we interweave convolutional layers with LSTM layers to mimic the transform process.

\begin{table}[H]
    \centering
    \begin{tabular}{|p{3cm}|p{2cm}|p{4cm}|}
        \hline
        Layer           &   Input Size      &   Parameters          \\  \hline
        (Input)         &                   &                       \\  \hline
        Conv1D          &   2$\times$2048   &   (3, 2, 1, circular) \\  \hline
        LeakyReLU       &   64$\times$1024  &   0.2                 \\  \hline
        LSTM            &   64$\times$1024  &                       \\  \hline
        Conv1D          &   64$\times$1024  &   (3, 2, 1, circular) \\  \hline
        LeakyReLU       &   128$\times$512  &   0.2                 \\  \hline
        LSTM            &   128$\times$512  &                       \\  \hline
        Conv1D          &   128$\times$512  &   (3, 2, 1, circular) \\  \hline
        LeakyReLU       &   256$\times$256  &   0.2                 \\  \hline
        LSTM            &   256$\times$256  &                       \\  \hline
        ConvTrans1D     &   256$\times$256  &   (3, 2, 1, 1)        \\  \hline
        Tanh            &   128$\times$512  &                       \\  \hline
        LSTM            &   128$\times$512  &                       \\  \hline
        ConvTrans1D     &   128$\times$512  &   (3, 2, 1, 1)        \\  \hline
        Tanh            &   64$\times$1024  &                       \\  \hline
        LSTM            &   64$\times$1024  &                       \\  \hline
        ConvTrans1D     &   64$\times$1024  &   (3, 2, 1, 1)        \\  \hline
        (Output)        &   2$\times$2048   &                       \\  \hline
    \end{tabular}
    \caption{\label{tab:arch}Neural network architecture for NFT. For inverse NFT, just swap all activation functions between "LeakyReLU" and "Tanh". The parameters for layers are Conv1D: (kernel size, stride, padding, padding mode), LeakyReLU: (negative slope), and ConvTrans1D: (kernel size, stride, padding, output padding).}
\end{table}
In addition to the four points discussed above, we also utilise the generative feature of the autoencoder structure to build the network~\cite{goodfellow_deep_2017}. An autoencoder learns two functions presented at both ends of the network by constructing a common representation of the two functions in a coded latent space.  \red{In our case, the latent space can be considered as a representation of the data encoded in the optical signal, while the two ends are the linear and nonlinear spectrum of that signal. The input side of the autoencoder tries to decode the data from the input spectrum, for example, a linear spectrum, and the output side of the autoencoder encodes the data using a different format, for example, into a nonlinear spectrum.} 
The network design is listed in Table \ref{tab:arch}. For the input and output, the complex signals are separated into real and imaginary parts. For instance, we choose an input size of $2\times2048$ for a 2048-element complex signal. The reset layer sizes are chosen objectively. Considering the symmetry in the auto-encoder architecture, the same network design can also be used for inverse NFT by using the nonlinear spectra as the input and the linear spectra as the output. In addition, we noticed a better performance after swapping the activation functions (LeakyReLU and Tanh) for iNFT.

To demonstrate the capability of the network for general use with various types of input signals, we consider the following in the preparation of the training dataset. The network takes a 2048-element single-precision complex-number array as the input. Assuming a signal generator with a sampling rate of 96~$GS/s$ is used, 2048 sampling points correspond to a time window of 21.33~$ns$ for each signal burst. Random bits of information were encoded using QAM and NFDM and then converted into the time domain through a fast inverse NFT algorithm~\cite{wahls_fnft:_2018} \red{(which guarantees there are no discrete spectral components in the signals)} with the following randomisation (uniformly distributed):
\begin{itemize}
    \item The nonlinear spectrum $Q(\lambda)$ is scaled with a random coefficient between 0.62 and 3.31 to randomise the signal power,
    \item The pulse width $T_0$ is randomly picked between 0.7 to 1.4 $ns$\footnote{\red{Note that the chosen values are only for demonstration purposes. For a broader pulse duration range, one can use the time dilation property of NFT: $\frac{1}{\left|a\right|}q(\frac{t}{a})\leftrightarrow Q(a\lambda)$~\cite{yousefi_information_2014}).}}, 
    \item A random constant phase between 0 to $\pi$ is added onto the nonlinear spectrum $Q(\lambda)$,
    \item A QAM format is randomly picked from 4, 16 and 64,
    \item A subcarrier number is randomly chosen from 32, 64 and 128.
    \item The subcarrier function is randomly chosen from 
        \begin{equation}
            w_n(\lambda) = \text{sinc}\left( \lambda T_0 - n\pi\right) \footnote{note: here $\text{sinc}(x)$ is defined as $\text{sin}(x) / x$}
            \text{,}
            \label{eq:cwsinc}
        \end{equation}
        or
        \begin{eqnarray}
            w_n(\lambda)=\frac{T_0}{8\sqrt{2\pi}}\left\{\text{erf}\left[\frac{1}{\sqrt{2}}\left(\lambda T_0-2\pi(2n-1)\right)\right] \right.\\ 
            \left.-\text{erf}\left[\frac{1}{\sqrt{2}}\left(\lambda T_0-2\pi(2n+1)\right)\right]\right\}\text{,}
            \label{eq:cwflattop}
        \end{eqnarray}
\end{itemize}

When training the network, the linear spectra and nonlinear spectra of the NFDM signals are used as inputs and labels, alternatively. Root-mean-square error (RMSE) is used for the loss function for training. For NFT, the nonlinear spectra are used as labels and normalised to their maximum modulus. For iNFT, the linear spectra are used as labels and normalised to their maximum modulus.

The network has a total of 1,104,258 trainable parameters and occupies roughly 4.2 MB of memory as single precision floating point numbers. The computation of a single forward pass consists of approximately \red{156 million FLOPs}.

\section{Results}
\begin{figure*}[h!]
    \centering
    \includegraphics[width=14cm]{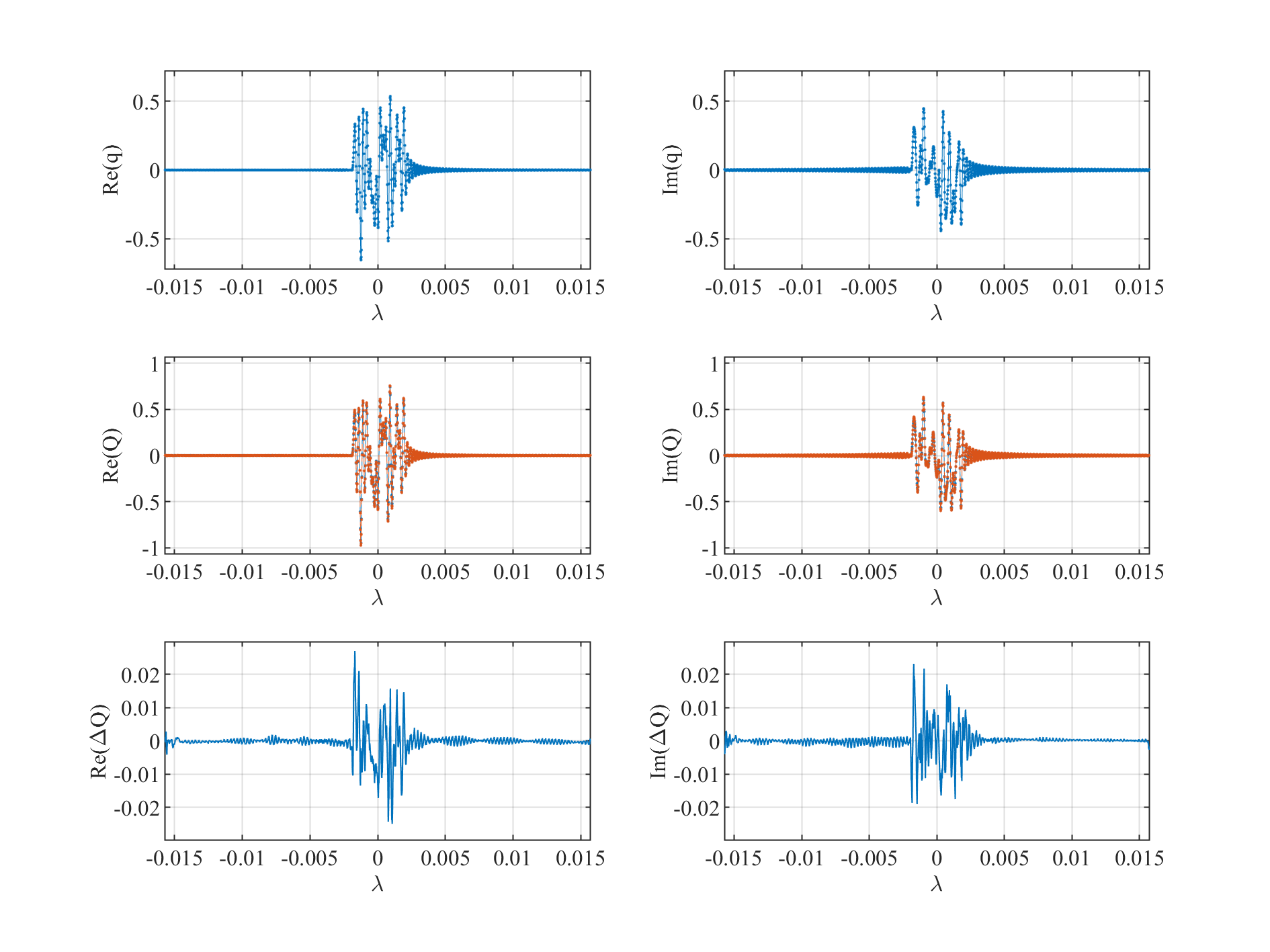}
    \caption{Example of an NFT of a 32-subcarrier low energy pulse. Top row: the real and imaginary part of the linear spectrum of the pulse. Middle row: the real and imaginary part of the nonlinear spectrum of the pulse obtained by the neural network (red), overlaid with the theoretical calculation (blue). Bottom row: the differences between the neural network and theoretical calculation.}
    \label{fig:32lp}
\end{figure*}

\begin{figure*}[h!]
    \centering
    \includegraphics[width=14cm]{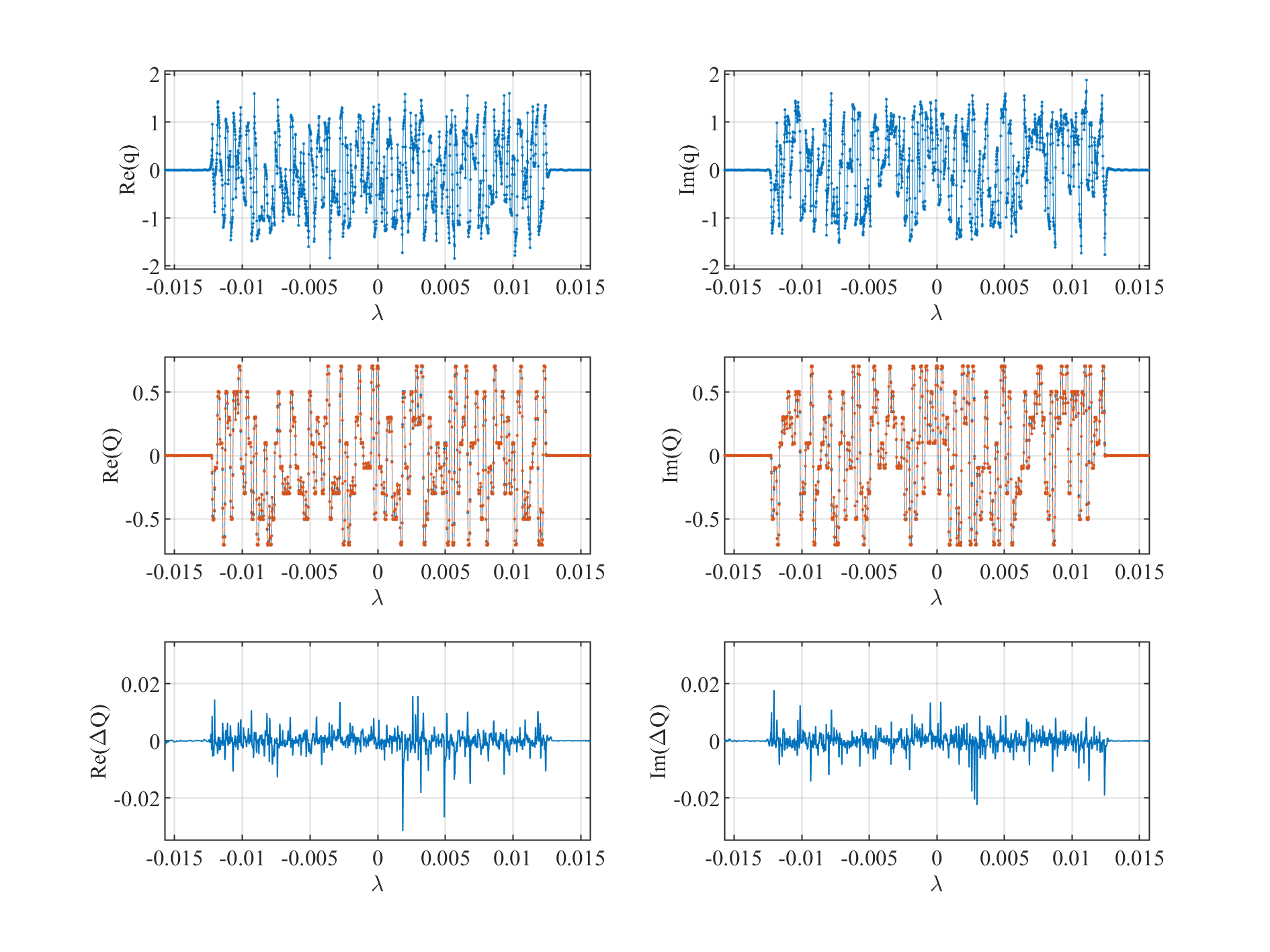}
    \caption{Example of an NFT of a 128-subcarrier high energy pulse. Top row: the real and imaginary part of the linear spectrum of the pulse. Middle row: the real and imaginary part of the nonlinear spectrum of the pulse obtained by the neural network (red), overlaid with the theoretical calculation (blue). Bottom row: the differences between the neural network and theoretical calculation.}
    \label{fig:128hp}
\end{figure*}

\begin{figure*}[h!]
    \centering
    \includegraphics[width=14cm]{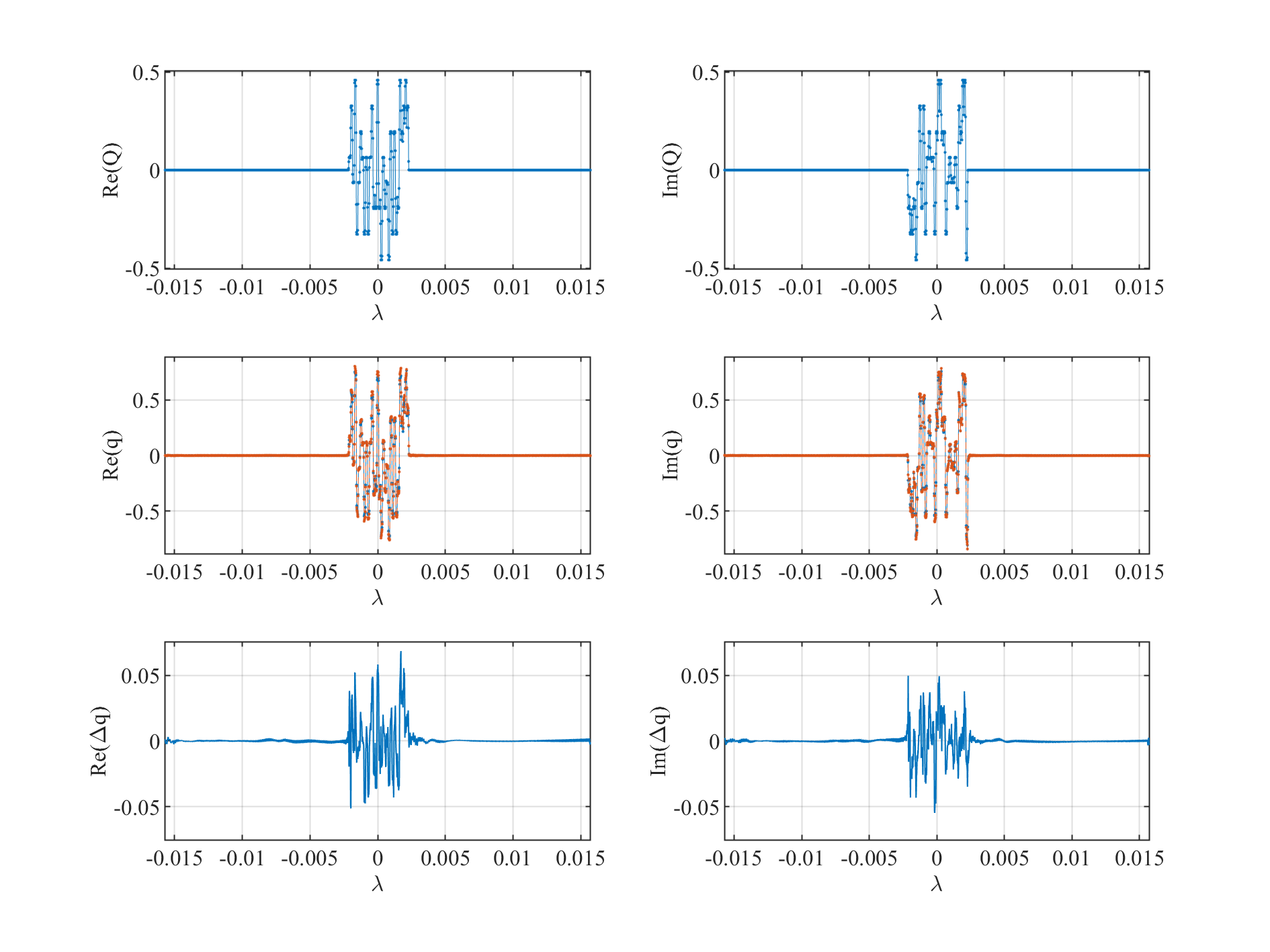}
    \caption{Example of an iNFT of a 32-subcarrier low energy pulse. Top row: the real and imaginary part of the nonlinear spectrum of the pulse. Middle row: the real and imaginary part of the linear spectrum of the pulse obtained by the neural network (red), overlaid with the theoretical calculation (blue). Bottom row: the differences between the neural network and theoretical calculation.}
    \label{fig:i32lp}
\end{figure*}

\begin{figure*}[h!]
    \centering
    \includegraphics[width=14cm]{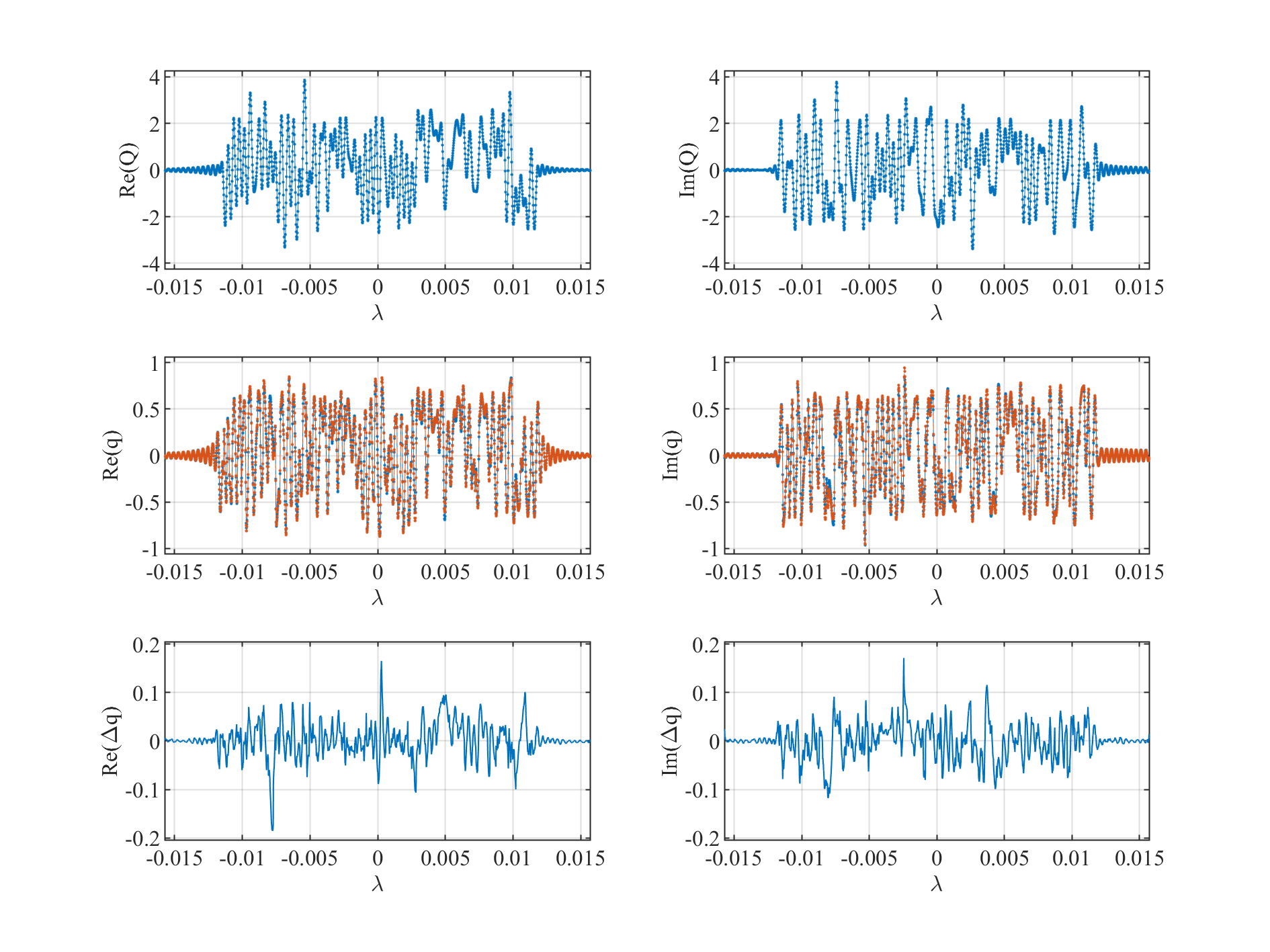}
    \caption{Example of an iNFT of a 128-subcarrier high energy pulse. Top row: the real and imaginary part of the nonlinear spectrum of the pulse. Middle row: the real and imaginary part of the linear spectrum of the pulse obtained by the neural network (red), overlaid with the theoretical calculation (blue). Bottom row: the differences between the neural network and theoretical calculation.}
    \label{fig:i128hp}
\end{figure*}

Both NFT and iNFT networks are trained on a dataset of 200,000 optical pulses using the ADAM optimizer\cite{kingma_adam_2017} and a learning rate of 3e-4 for 200 epochs. The validations are performed on a separate dataset of 100,000 pulses for statistical results. The RMSEs of the output increase with the pulse energy for both NFT and iNFT, while the RMSE of iNFT is about 3$\sim$9 times larger than the RMSE of NFT at every energy level. The RMSEs for a range of pulse energy are listed in Table \ref{tab:rmse}, 
\newcommand\pwidth{0.8cm}
\begin{table*}[!h]
    \centering
    \begin{tabular}{ |p{1.5cm}|p{\pwidth}|p{\pwidth}|p{\pwidth}|p{\pwidth}|p{\pwidth}|p{\pwidth}|p{\pwidth}|p{\pwidth}|p{\pwidth}|p{\pwidth}|p{\pwidth}|  }
        \hline
        \multicolumn{2}{|c|}{Energy (pJ)} & 0.035 & 0.055 & 0.091 & 0.151 & 0.251 & 0.413 & 0.682 & 1.131 & 1.848 & 2.876 \\ \hline
        \multirow{2}{2.3cm}{RMSE ($\times10^{-2}$)} & NFT & 0.388 & 0.306 & 0.292 & 0.329 & 0.401 & 0.519 & 0.656 & 0.844 & 1.089 & 1.098 \\ \cline{2-12}
        & iNFT & 1.163 & 1.859 & 2.456 & 2.972 & 3.404 & 3.796 & 4.142 & 4.428 & 4.960 & 5.401 \\ \hline
        \pinkrow & NN & 0.216 & 0.367 & 0.246 & 0.388 & 1.282 & 3.087 & 3.678 & 4.692 & 6.246 & 3.366 \\ \cline{2-12}
        \pinkrow\multirow{-2}{2.3cm}{BER\hphantom{X} ($\times10^{-2}$)} & FNFT& 0 & 0 & 0 & 0.031 & 0.329 & 0.626 & 1.899 & 6.737 & 14.53 & 7.296 \\ \hline
    \end{tabular}
    \caption{\label{tab:rmse}\red{RMSE of forward and inverse NFT networks, and the BER of the networks and FNFT library at different pulse energy levels of the whole testing dataset.}}
\end{table*}
where examples of pulses on the low energy end of the validation dataset can be found in Figs.~\ref{fig:32lp} and~\ref{fig:i32lp} while examples of pulses on the high energy end of the validation dataset can be found in Figs.~\ref{fig:128hp} and~\ref{fig:i128hp}. Since the training labels for both NFT and iNFT cases are normalised to their maximum modulus, the RMSEs are close to the relative errors in the output spectra. On average, NFT spectra have an error of approximately 0.5\%, while iNFT spectra have an error of approximately 3\%. 

\red{To compare the performance of the neural networks with the original fast NFT algorithm, a back-to-back test is used. The modulated nonlinear spectra are passed through the iNFT network first followed directly by the NFT network, then the output spectra are decoded to obtain encoded bits, and the bit error rates (BER) are calculated. Similar calculations are also performed using the fast inverse NFT and fast NFT library (FNFT)~\cite{wahls_fnft:_2018}. The BERs from both cases are compared and listed also in Table~\ref{tab:rmse}.  In general, the BER also increases with pulse energy. For the neural networks, the change of BER is relatively small (an order of magnitude). For the FNFT library, although its BER is smaller than the neural networks in the low energy range, the BER increases drastically with pulse energy. It exceeds the neural networks around 1.131 pJ level. It is important to point out that the linear Fourier spectra used as the label for the training of our neural networks are calculated using the FNFT library. The numerical errors produced by the FNFT library are carried into the neural networks through training. However, the neural networks seem to be tuned to balance the errors among the whole range of data, which can be one of the reasons a better performance is observed in the neural networks in the high energy range. Overall, the total amount of error bits with the neural networks and the FNFT library are in the same order with 246891 error bits for the neural networks and 435772 error bits for the FNFT library among a total of 7476960 bits of the test dataset.}

Datasets used consist of a mixed number of subcarriers. The signal bursts with more subcarriers tend to carry more energy. By looking at the RMSE of the validation dataset for 32, 64, and 128 subcarriers separately (Table.~\ref{tab:rmsesub}), one can notice that pulses with more subcarriers have higher RMSE and the RMSE increases approximately linearly with pulse energy. \red{The same trend can be observed in the BERs of the back-to-back test for different subcarriers, while the BERs of the neural networks remain at a similar level, the BERs of the FNFT library increase drastically. The neural networks outperform the FNFT library for the 128-subcarrier case.}
\begin{table}[h!]
    \centering
    \begin{tabular}{|p{1.3cm}|p{1.4cm}|p{1.7cm}|p{1.7cm}|p{1.7cm}|p{1.7cm}|}
    \hline
    Sub- carriers & Energy (pJ) & RMSE (NFT) & RMSE (iNFT) & \pink{BER\hphantom{X} (NN)} & \pink{BER (FNFT)}\\ \hline
    32 & 0.32 & 3.76e-3 & 1.34e-2 & \pink{3.06e-02} & \pink{0} \\ \hline
    64 & 0.64 & 4.90e-3 & 1.88e-2 & \pink{3.20e-02} & \pink{2.45e-03} \\ \hline
    128 & 1.27 & 6.73e-3 & 2.73e-2 & \pink{3.41e-02} & \pink{1.00e-01} \\ \hline
    \end{tabular}
    \caption{\label{tab:rmsesub}\red{RMSE of NFT and iNFT networks, and BER of the networks (NN) and FNFT library for 32, 64, and 128 subcarriers}}
\end{table}

In terms of different QAM formats (Table.~\ref{tab:rmseqam}), one can notice that 4-QAM format pulses have the highest pulse energy but do not always have the highest RMSE. The RMSEs of 16-QAM and 64-QAM are similar for NFT, while 64-QAM is slightly lower than 4- and 16-QAM for iNFT. \red{A similar trend can also be observed in the BERs of the back-to-back test for different QAM. In this case, the BERs of the neural networks and FNFT library are rather close with the neural networks performing slightly better.}
\begin{table}[h!]
    \centering
    \begin{tabular}{|p{1.3cm}|p{1.4cm}|p{1.7cm}|p{1.7cm}|p{1.7cm}|p{1.7cm}|}
    \hline
    QAM & Energy (pJ) & RMSE (NFT) & RMSE (iNFT) & \pink{BER\hphantom{X} (NN)} & \pink{BER (FNFT)}\\ \hline
    4 & 0.98 & 3.89e-3 & 2.16e-2 & \pink{2.50e-04} & \pink{4.04e-04} \\ \hline
    16 & 0.67 & 5.78e-3 & 1.97e-2 & \pink{1.69e-02} & \pink{2.25e-02} \\ \hline
    64 & 0.57 & 5.75e-3 & 1.83e-2 & \pink{7.95e-02} & \pink{8.05e-02} \\ \hline
    \end{tabular}
    \caption{\label{tab:rmseqam}\red{RMSE of NFT and iNFT networks, and BER of the networks and FNFT library for 4-, 16- and 64-QAM.}}
\end{table}

A few spectral comparison examples are shown in Fig.~\ref{fig:32lp} to \ref{fig:i128hp}. Fig.~\ref{fig:32lp} and \ref{fig:128hp} show the input and output spectra of the network for NFT. (Note: To make comparison easy, the x-axes for both linear and nonlinear spectra are labelled with nonlinear frequency $\lambda$. The linear angular frequency $\omega$ is just twice the nonlinear frequency $\lambda$.) Fig.~\ref{fig:32lp} shows a case with 32 subcarriers at low pulse energy, while Fig.~\ref{fig:128hp} shows a case with 128 subcarriers at high pulse energy. The top rows of the figures are the real (left) and imaginary (right) parts of the input linear spectra, the middle rows are the nonlinear spectra generated by the neural network (red-dotted curves) overlaid with the calculated spectra using the Fast NFT (FNFT) algorithm (blue-dotted curves)~\cite{wahls_fast_2015-1}. The bottom rows are the differences between the neural network outputs and the calculated spectra. For NFT, both cases show good agreements with RMSEs of 0.0033 and 0.0027 for 32-subcarrier and 128-subcarrier pulses, respectively.

Fig.~\ref{fig:i32lp} and \ref{fig:i128hp} show the input and output spectra of the network for iNFT with 32 subcarriers at low pulse energy and 128 subcarriers at high pulse energy, respectively. The top row of the figures is the real and imaginary parts of the input nonlinear spectrum. The middle row is the counterpart of the output linear spectrum (red-dotted curves) overlaid with the calculated linear spectra using the Fast iNFT algorithm (blue-dotted curves)~\cite{vaibhav_fast_2018}. The bottom rows are the differences between the neural network outputs and the calculated spectra. For iNFT, the RMSEs of the neural network for 32-subcarrier low energy pulse and 128-subcarrier high energy pulse are 0.0088 and 0.035, respectively.

For both NFT and iNFT cases, when pulse energy is low, the nonlinear Fourier transform converges to the linear Fourier transform, hence, the linear and nonlinear spectra in Fig.~\ref{fig:32lp} and \ref{fig:i32lp} are nearly identical. At high pulse energy, however, the nonlinear spectrum of the pulse is significantly different from the linear spectrum due to nonlinear cross-talks between linear frequencies. The ability to perform different transformations between linear and nonlinear spectra based on the pulse energy has never been demonstrated in any previous NFT neural network designs. The same with the ability to perform transform on pulses with various numbers of sub-carriers, temporal widths, spectral bandwidth, and modulation formats at the same time. The proposed network is indeed attempting to perform NFT and iNFT calculations instead of simple mapping between a given set of inputs and outputs. To further demonstrate this point, the network was applied to a few pulses that are completely different from the training dataset.

\begin{figure}[h!]
    \centering
    \includegraphics[width=10cm]{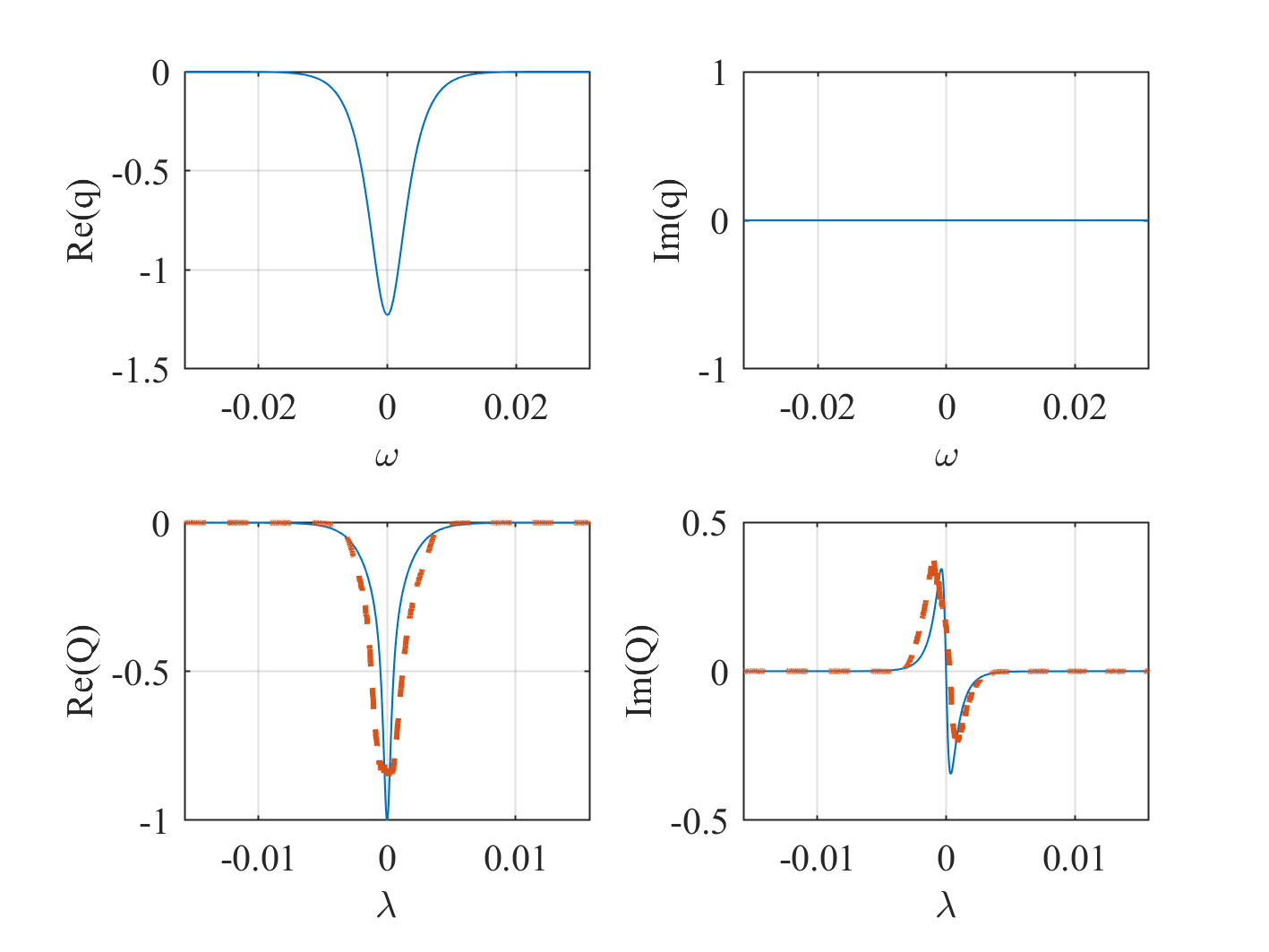}
    \caption{The real and imaginary part of the linear spectrum (top row) of an $A\text{sech}(t)$ pulse and its nonlinear spectrum (bottom row) generated by the neural network, overlaid with the theoretical values.}
    \label{fig:infersech}
\end{figure}

\begin{figure}[h!]
    \centering
    \includegraphics[width=10cm]{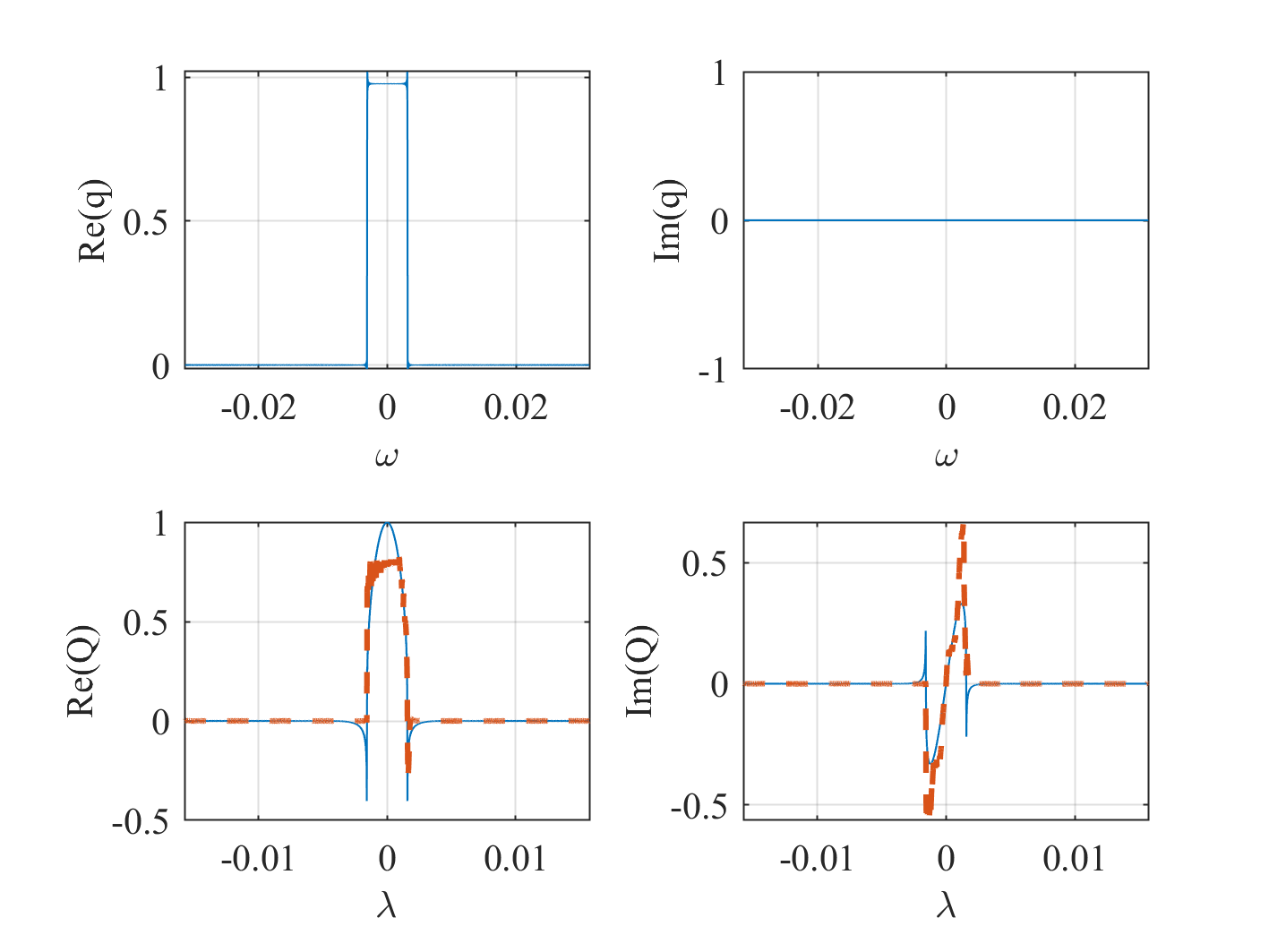}
    \caption{The real and imaginary part of the linear spectrum (top row) of a $-A\text{sinc}(t)$ pulse and its nonlinear spectrum (bottom row) generated by the neural network, overlaid with the theoretical values.}
    \label{fig:infersinc}
\end{figure}

Fig.~\ref{fig:infersech} and \ref{fig:infersinc} shows the linear and nonlinear spectra of two pulses in the form of $A\text{sech}(t)$ and $-A\text{sinc}(t)$, respectively. The top row of the figures is the real and imaginary part of the pulse's linear spectrum. The bottom row of the figures is the real and imaginary part of the pulse's continuous nonlinear spectrum obtained through the network (red-dashed curves) overlaid with the theoretical values calculated using the FNFT algorithm (blue curves). For both pulses, their linear Fourier transform only contains real components, however, the network is able to roughly generate both real and imaginary parts of the nonlinear spectra with \red{RMSEs of 0.08 for the $A\text{sech}(t)$ and 0.04 for $A\text{sinc}(t)$}. Note that the pulses shown here do not have discrete eigenvalues.

\begin{figure}[h!]
    \centering
    \includegraphics[width=10cm]{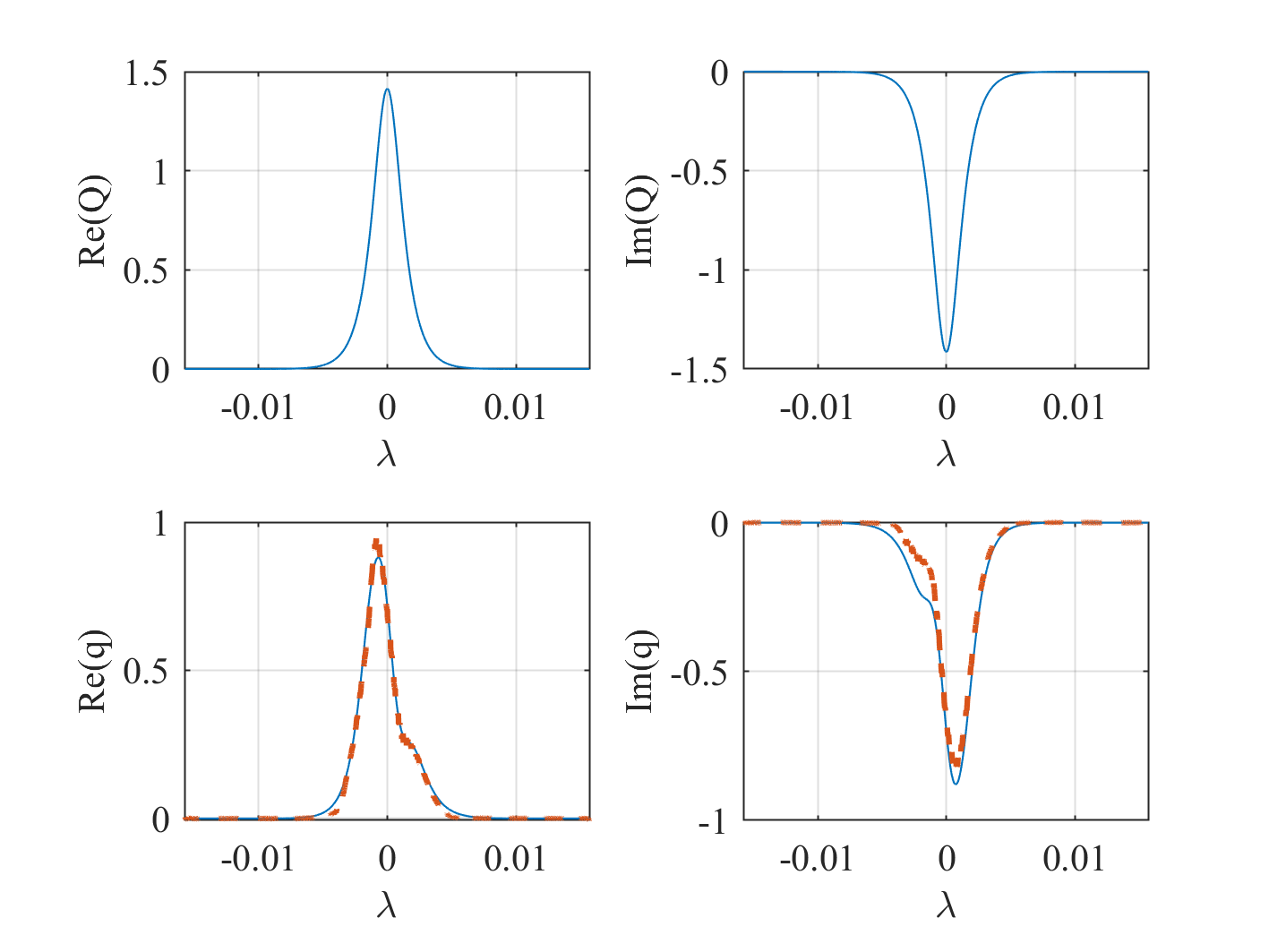}
    \caption{The real and imaginary part of the nonlinear spectrum (top row) of a $A\text{sech}(\lambda)e^{-i\pi/4}$ pulse and its linear spectrum (bottom row) generated by the neural network, overlaid with the theoretical values.}
    \label{fig:inferisech}
\end{figure}

\begin{figure}[h!]
    \centering
    \includegraphics[width=10cm]{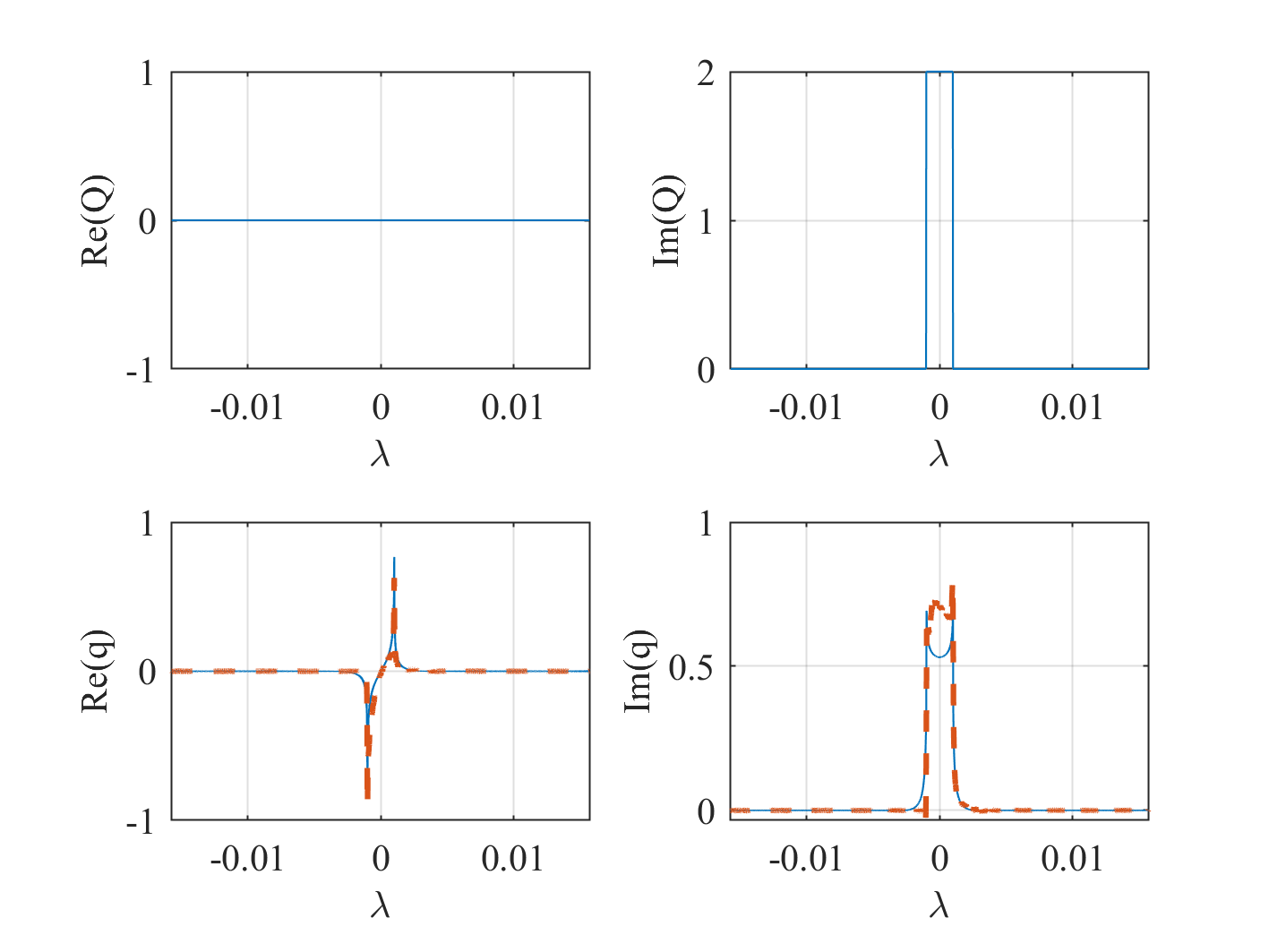}
    \caption{The real and imaginary part of the nonlinear spectrum (top row) of a $A\text{rect}(\lambda)$ pulse and its linear spectrum (bottom row) generated by the neural network, overlaid with the theoretical values.}
    \label{fig:inferirect}
\end{figure}

Fig.~\ref{fig:inferisech} and \ref{fig:inferirect} shows the iNFT of two pulse spectra in the form of $A\text{sech}(\lambda)e^{-i\pi/4}$ and $iA\text{rect}(\lambda)$, respectively. The top row of the figures is the real and imaginary part of the pulse's nonlinear spectrum. The bottom row of the figures is the real and imaginary part of the pulse's linear spectrum obtained through the network (red dashed curves) overlaid with the theoretical values (blue curves). The spectrum in Fig.~\ref{fig:inferisech} contains both real and imaginary components, whilst the spectrum in Fig.~\ref{fig:inferirect} is purely imaginary. For both cases, the network can also generate the corresponding linear spectra with \red{RMSEs of 0.027 and 0.035, respectively}.

The network was trained with NFDM-QAM signals, but the examples shown in Figs.~\ref{fig:infersech} to~\ref{fig:inferirect} used general arbitrary pulses. Although the prediction accuracy for these arbitrary pulses is lower than the training pulse types, the ability of the network still being able to generate a rough shape of the linear and nonlinear spectra indicates that the proposed network architecture is capable of performing general NFT and iNFT. The proposed network can be tuned to specific NFT and iNFT applications with decent accuracy when given the right training data, such as the NFDM-QAM signals used in this work. 

\section{Conclusion}
\label{sec:conclusion}
In this paper, a neural network is proposed for both forward and inverse continuous nonlinear Fourier transforms. 
The network is demonstrated with a random mix of NFDM-QAM signals (different carrier wave functions, pulse bandwidth, pulse width, and encoding formats) with single-precision floating point numbers. The network is able to achieve an RMSE of $5\times10^{-3}$ for NFT and an RMSE of $3\times10^{-2}$ for iNFT for a range of pulse energy used, where on the low energy end, the nonlinear spectra of the pulses converge to their linear spectra while on the high energy end, the nonlinear spectra of the pulses significant differ from their linear spectra. We further showed that the trained network can be used to perform general NFT and iNFT on arbitrary pulses beyond the training pulse types, although the accuracy is dropped. \red{In terms of BER in a back-to-back test, the performance of the neural networks is better than the FNFT library for high-energy or large subcarrier signals. On average, the performances of the two are similar.}

\red{The computational complexity of our neural network can be roughly calculated using the following steps. For the convolutional layers, every kernel matrix multiplication requires $\frac{NK}{2}$ multiplications and $\frac{N(K-1)}{2}$ additions for a stride size of 2, where $N$ is the input data length, and $K$ is the kernel size. For each convolutional layer, we have $F$ input features and $G$ output features. Hence, the total number of FLOPs for a convolutional layer is $\sim \mathcal{O}(NKFG)$. We have 6 convolutional layers, among which, 3 of them are transpose convolutional layers. We approximate the FLOPs of the transposed convolutional layers as convolutional layers. Therefore, the total FLOPs of all convolutional layers are roughly:
$$c_\text{conv}=\left[2048\times3\times2\times64 + 1024\times3\times64\times128 + 512\times3\times128\times256\right]\times2\approx1.53\times10^8$$
For each LSTM layer, we have 4 gates. Each gate has $2(N F + F^2)$ multiplications and additions. For each layer, we also have $3F$ FLOPs for computing cell states and $2F$ FLOPs for computing hidden states, which can be ignored. Hence, the total number of FLOPs for an LSTM layer is  $\sim\mathcal{O}(8 F (N + F))$. The total FLOPs for all 5 LSTM layers are roughly: 
$$
c_\text{LSTM}=\left[8\times64\times(1024+64) + 8\times128\times(512+128)\right]\times2 + 8\times256\times(256+256) \approx 3.5\times10^6
$$
The overall FLOPs of the neural network are around 156 million. For comparison, the complexity of FNFT with the same input size is about 20 million~\cite{chimmalgi_fast_2019}, and the complexity for the previous work is around 80 million assuming the number of additions is the same as multiplications~\cite{sedov_neural_2021-1}.  However, different from previous work~\cite{sedov_neural_2021-1}, our network is capable of processing signals with different power levels, different pulse durations, different pulse bandwidths, different modulation formats, and random phases at the same time. 

Note that, the network presented in this work is not optimized for efficiency and hence it has room for improvement.} 
The number of Conv-LSTM and ConvTrans-LSTM layer stacks and the number of features for the convolutional layers (64, 128, 256 in the current example) can be \red{tuned using a Bayesian optimization process~\cite{morelli_bayesian_2021}.}
One can also introduce other design ideas such as adding residual structures to improve the performance. Nevertheless, we hope the potential of this network design can promote the development of NFT-related devices for further research and application advancement.


\section*{Acknowledgements}
This research was supported fully by the Australian Government through the Australian Research Council (DP190102896).

\bibliographystyle{IEEEtran}


\end{document}